\documentstyle[aps]{revtex} 

\input epsf

\begin{document}
\twocolumn[
\title{Perturbative Duality in the Resonance Spin Structure Functions}

\author{
Carl E. Carlson
}
\address{
Physics Department, College of William and Mary,
Williamsburg, VA 23187
}

\author{
Nimai C. Mukhopadhyay
}
\address{
Department of Physics, Applied Physics, and Astronomy,  Rensselaer
Polytechnical Institute,  Troy,  NY 12180-3590
}
\date{31 Decembar 1997}
\maketitle
\parshape=1 0.75in 5.5in \indent
{\small We investigate the relations between the spin structure 
functions in the scaling and resonance regions.  We examine the possible
duality between the two,  and draw inferences for the
behavior of the asymmetry $A_1$ at large $x$. Finally, we
point out the importance of additional polarized structure function data
in the resonance region in terms of testing the hysteresis of perturbative
physics.
 
\vglue -20pt}


\widetext \vglue -5.9cm  \hfill RPI-97-N122; WM-97-116; hep-ph/9801205
\vglue 5.4cm \narrowtext
\pacs{13.88+e, 14.20.Gk, 13.60.Hb, 12.40.N, 12.38.Bx}
]




Duality is a relation between the deep inelastic scattering region and
the resonance region in lepton hadron scattering~\cite{bg}.   It is
a manifestation of the fact that the single quark reaction rate
determines the scale of the reaction rate for the entire process down to
remarkably low energies and momentum transfers.  In the resonance region,
final state interactions are crucial and all quarks must be involved in
the reaction.  Nevertheless, the overall reaction rate is still
determined by the single quark reaction rate---provided we average over
regions comparable to the widths of the resonances.   So far, these
inclusive/exclusive connections have been seen to work for the
measured~\cite{s,bosted} unpolarized structure functions~\cite{bg,cm}. 
Precious little experimental information is available for the r\^ole of
duality for the polarized structure functions, although efforts in this
direction are beginning to bear fruit~\cite{abe}.

Here we investigate the relevance and consequences of duality for the
polarized structure functions.  We shall show that perturbative QCD
arguments lead to an inclusive/exclusive relationship in the polarized
case akin to that in the unpolarized one.  Also, we shall discuss
perturbative expectations for resonance contributions in the high $x$
limit of the polarization asymmetry.  Finally, we note interesting
features of the present data and point out what can be learned from
additional polarized structure function data in the resonance region at
higher $Q^2$.




We begin with some definitions and kinematic relations.  For deep
inelastic scattering,
$e+p \rightarrow e + X$, the structure functions $W_i$ and $G_i$ are
defined by
\begin{eqnarray}
W_{\mu\nu} &=& {1\over 4\pi m_N} \sum_X (2\pi)^4 \delta^4(q+p-p_X)
   \nonumber \\  &\times&
   \langle ps | j_\mu(0) | X \rangle \langle X | j_\nu(0) | ps \rangle
   \nonumber \\
  &=& -g_{\mu\nu} W_1 + {1\over m_N^2} p_\mu p_\nu W_2
   + {i\over m_N} \epsilon_{\mu\nu\lambda\sigma} q^\lambda \times
   \nonumber \\  &\times&
   \left[ s^\sigma G_1  + {1\over m_N^2} 
        (p\cdot q s^\sigma - s\cdot q p^\sigma) G_2 \right]  ,
\end{eqnarray}

\noindent where $s$ is the spin vector of the nucleon target and
satisfies $s \cdot p =0$ and $s^2 = -1$, $p$ is the momentum of the
target nucleon, and $q$ is the momentum of the incoming virtual photon. 
Often, the structure functions are replaced by
\begin{eqnarray}
\nu W_2 &=& F_2 \quad , \quad m_N W_1 = F_1 \quad , \nonumber \\
   \nu G_1 &=& g_1 \quad , \quad {\nu^2\over m_N} G_2 = g_2
\end{eqnarray}

\noindent where it is expected that $F_i$ and $g_i$ depend just on 
$x \equiv Q^2/2 m_N \nu$ (where $Q^2 = - q^2$) in the scaling region, up to
logarithmic corrections.  

The exclusive process $e + p \rightarrow e + R$, where $R$ stands for
the final baryon, a resonance or the nucleon (in the case of elastic
scattering),  is described by the helicity amplitudes,
\begin{equation}
G_m = {1\over 2m_N} \langle R, \lambda^\prime = m - {1\over 2} |
   \epsilon_\mu^{(m)} \cdot j^\mu(0) | N, \lambda = {1\over 2} \rangle .
\end{equation}

\noindent The photon polarization vectors are
\begin{eqnarray}
\epsilon^{(\pm)} &=& (0,\mp 1, -i,0)/ \sqrt{2} , \nonumber \\
\epsilon^{(0)} &=& {1\over Q} (|\vec q|, 0, 0, \nu),
\end{eqnarray}

\noindent with $q = (\nu,0,0,|\vec q|)$.

Note that the helicity of the final baryon is
\begin{eqnarray}
{1\over 2} \  &{\rm for}& \  G_+ , \nonumber \\
-{1\over 2} \ &{\rm for}& \ G_0 , \nonumber \\
-{3\over 2} \  &{\rm for}& \ G_-.
\end{eqnarray}

\noindent Thus if the final baryon has spin 1/2, $G_-$ must be absent.
For elastic scattering, the helicity amplitudes are related to well-known
form factors by
\begin{eqnarray}
G_+ &=& {Q\over m_N \sqrt{2}} G_M , \nonumber \\
G_0 &=& G_E .
\end{eqnarray}
For the non-elastic case, one often uses the amplitudes
\begin{eqnarray}
\left| A_{1/2,3/2} \right|
    = e \sqrt{m_N \over m_R^2 - m_N^2 } \left| G_{+,-} \right| \nonumber
\end{eqnarray}
where $e$ is the proton charge (0.3028 \ldots).

For a single sharp resonance $R$ the relations between the structure
functions and the helicity amplitudes are:
\begin{eqnarray}
F_1 &=& m_N^2 \delta(W^2-m_R^2) \left[ |G_+|^2 + |G_-|^2 \right] ,
               \nonumber \\
\left( 1+{\nu^2\over Q^2} \right) F_2 &=& m_N \nu \delta(W^2-m_R^2)
               \nonumber \\
    &\quad&  \times \left[ |G_+|^2 + 2|G_0|^2 + |G_-|^2 \right]  ,
               \nonumber \\[1.5ex]
\left( 1+{Q^2\over \nu^2} \right) g_1 &=& m_N^2 \delta(W^2-m_R^2)
               \nonumber \\
    \times \biggl[ |G_+|^2 &-& |G_-|^2 
        + (-1)^{s_R-{1\over 2}} \eta_R   {Q \sqrt{2} \over \nu} 
             G_0^* G_+   \biggr],
               \nonumber \\[2ex]
\left( 1+{Q^2\over \nu^2} \right) g_2 &=& - m_N^2 \delta(W^2-m_R^2)
               \nonumber \\
   \times   \biggl[ |G_+|^2 &-& |G_-|^2 
          - (-1)^{s_R-{1\over 2}} \eta_R   {Q \sqrt{2} \over \nu} 
               G_0^* G_+  \biggr] ,
\end{eqnarray}

\noindent where $W^2 \equiv (p+q)^2$, the total hadronic mass
squared, and $s_R$ and $\eta_R$ are the spin and parity of the
resonance.  The delta function for the sharp resonance can be most simply
approximated by
\begin{eqnarray}
\delta(W^2-m_N^2) &\approx& {1\over 2m_R} {\Gamma_R / 2\pi \over
         (W-m_R)^2 + \Gamma_R^2/4 }    \nonumber \\
     &\ \stackrel{peak}{\longrightarrow}& \ {1\over \pi m_R \Gamma_R} ,
\end{eqnarray}

\noindent with $\Gamma_R$ being the width of the resonance.




Let us now discuss the scaling properties of $G_{\pm,0}$ and of
$g_{1,2}$.  The resonance contributions to the structure functions fall
with increasing $Q^2$ and also move to progressively higher $x$,
approaching
$x=1$.  They thus may be described as falling with $x$, at a certain rate. 
We wish to determine if the fall-off rate is the same as that in the deep
inelastic region as $x \rightarrow 1$, but at much higher $Q^2$.   This is
already known to be true for the spin independent structure functions,
and the phenomenon is known as Bloom-Gilman duality~\cite{bg}.  Hence,
we will concentrate our attention on the spin dependent structure
functions.  

The counting rules~\cite{bf} give the following behavior at high $Q^2$ for
the helicity amplitudes (modulo logarithms)~\cite{cm}:
\begin{eqnarray}
G_+ &=& g_+ / Q^3 \ , \qquad G_0 = (m_N)g_0/Q^4 , \nonumber \\
   G_- &=& (m_N)^2 g_-/Q^5 \ ,
\end{eqnarray}

\noindent where $g_{\pm,0}$ are constants, real in leading Born order, and
the mass factors are put in purely for dimensional reasons.  This allows
us to find the resonance contribution to $g_1$ at the resonance peak and
at high $Q^2$ to be,
\begin{equation}
g_1 = {m_N^2\over \pi m_R \Gamma_R} {g_+^2\over Q^6}
   =  {m_N^2\over \pi m_R \Gamma_R} {g_+^2\over (m_R^2-m_N^2)^3} (1-x)^3 .
\end{equation}

\noindent The second result requires
\begin{equation}
{1\over Q^2} = {1\over W^2 - m_N^2} {1-x\over x} 
      \approx  {1\over m_R^2 - m_N^2} (1-x)
\end{equation}

\noindent for $x \rightarrow 1$ and $ W\approx m_R$.  Similarly,
\begin{equation}
g_2 = - {m_N^2\over \pi m_R \Gamma_R} 
  {(1-x)^3 \over (m_R^2-m_N^2)^3} 
       g_+(g_+ - {\eta_R (-1)^{s_R-{1\over 2}}\over \sqrt{2}}g_0)  ,
\end{equation}

\noindent for high $Q^2$.   It is interesting to note that the high $Q^2$
resonance contributions to the polarized structure functions can be
inferred from unpolarized structure function measurements since $g_+$
and $g_0$ can be separately obtained from transverse and longitudinal
scattering.  If $g_2$ is small, as seems to be indicated~\cite{g2}, then
there must be a relation between the transverse and longitudinal
resonance form factors, {\it viz.},
\begin{equation}
\sqrt{2} \, g_+ \approx \eta_R (-1)^{s_R-{1\over 2}} g_0  .
\end{equation}

In the deep inelastic region, the spin structure function $g_1$ is
related to the quark distributions as a manner similar to $F_1$ except
for one sign,
\begin{eqnarray}  \label{structurefcn}
g_1 = {1\over 2} \sum e_q^2 \left( q_\uparrow(x,Q^2) 
      - q_\downarrow(x,Q^2) \right),
   \nonumber \\
F_1 = {1\over 2} \sum e_q^2 \left( q_\uparrow(x,Q^2) 
      + q_\downarrow(x,Q^2) \right).
\end{eqnarray}

\noindent The $q_{\uparrow,\downarrow}$ are the quark distributions for
quark helicities parallel or antiparallel to the parent nucleon
polarization.  Perturbative QCD dictates that $q_\uparrow$ dominates
as $x \rightarrow 1$~\cite{fj}.  If so, the
$x \rightarrow 1$ behavior  will be the same for both functions.  Even
if pQCD did not work for the $x \rightarrow 1$ limit, it would require
a remarkable cancellation to make the high $x$ behavior different for
$F_1$ and $g_1$.   Since the behavior of $F_1$ is well-known, we can
conclude
\begin{equation}
\lim_{x \rightarrow 1} g_1(x) \propto (1-x)^3
\end{equation}

\noindent in the deep inelastic region.  This is the same as the
contribution from the resonance region.   One part of the duality between
the resonance and deep inelastic regions is thus established.

It is less clear what the deep inelastic result should be for $g_2$. 
There is no unique parton model formula for it.  However the twist-two
form of the Wandzura-Wilczek relation~\cite{ww},
\begin{equation}
g_2(x) = -g_1(x) + \int_x^1 dx^\prime  \, {g_1(x^\prime)\over x^\prime},
\end{equation}

\noindent leads to the result that
\begin{equation}
\lim_{x \rightarrow 1} g_2(x) = - g_1(x) \propto (1-x)^3  .
\end{equation}

\noindent Thus the scaling part of the duality is established for
$g_2$ also.




Now we shall further examine the $x \rightarrow 1$ behavior of $g_1$, or
of the photon asymmetry $A_1$,
\begin{eqnarray}
A_1 \equiv {\sigma_{1/2}-\sigma_{3/2} \over \sigma_{1/2}+\sigma_{3/2} }
   =  {g_1 - {Q^2 \over \nu^2} g_2 \over F_1},
\end{eqnarray}

\noindent where the cross sections are for photon absorption with initial
state spin projections of 1/2 and 3/2.  For a resonance,
\begin{equation}
A_1 = {|G_+|^2 - |G_-|^2 \over |G_+|^2 + |G_-|^2 }  .
\end{equation}

For the elastic point, $x=1$, there is only the nucleon and $G_- = 0$ so
that rigorously
\begin{equation}
A_1(x=1) = 1.
\end{equation}

For a single resonance of spin 1/2, the same is true.  Even for spin 3/2
and higher resonances,  the scaling rules tell us that $|G_+| >> |G_-|$
at high $Q^2$, so that $A_1 \rightarrow 1$ as $x \rightarrow 1$.  If the
only backgrounds under a given resonance are due to tails of other
resonances, then the same rule still applies,
\begin{equation}
\lim_{Q^2 \rightarrow \infty} A_1 = 1  \qquad 
     {\rm if\ only\ resonant\ background}.
\end{equation}

\noindent However, if the non-resonant background is dominated by
Born terms~\cite{dmw}, we can get $A_1 \rightarrow 1$ anyway.  In the
resonance region the $t$-channel and $u$-channel diagrams have
propagators that suppress their contributions at high $Q^2$, leaving the
$s$-channel diagram which has only $\sigma_{1/2}$.  (Purely as
an aside,  a dominant $t$-channel Born term is needed for certain
measurements of the pion form factor,  and this can indeed happen even
at high $Q^2$, but only if the final hadronic state is well outside the
resonance region.)   Note that the isospin of the resonance plays no
r\^ole in the above considerations.

The $x \rightarrow 1$ behavior of $A_1$ in the scaling region can be
got from the ratio of the two eqns.~(\ref{structurefcn}). We quote the
results for pQCD, where the 
$x \rightarrow 1$ results for the polarized quark distributions are
mentioned above);  for SU(6), where no distinction is made
between the distributions of differently polarized quarks; and for a
number of modern suggestions for the polarized quark
distributions~\cite{polarize}.   One has
\begin{eqnarray}
\lim_{x \rightarrow 1} A_1 = \left\{
  \begin{array} {cl}
   1 & \quad {\rm pQCD,\ or\ Soffer\ {\it et\ al.}~\cite{polarize},}
                                     \\[1ex]
   {5\over 9} & \quad {\rm SU(6),}   \\[1ex]
   0.75   &   \quad {\rm GS,\ version\ B~\cite{polarize}}   \\[1ex]
   0.66   &   \quad {\rm GRSV\ (``standard''\ NLO)~\cite{polarize}}
  \end{array}   \right.
\end{eqnarray}
The last two are given at their respective benchmark $Q^2$.




Let us look at the present relevant data on the polarized structure
functions in the resonance region.  There is a paucity of such data.  The
SLAC measurements, from the E143 collaboration recently~\cite{abe} (and
from earlier data with larger error bars~\cite{baum}), do cover
$W^2 < 5$ GeV$^2$ at $Q^2 \approx 0.5$ and $1.2$ GeV$^2$.   These $Q^2$
are still too low for a duality test, since duality is not
working at these $Q^2$ in the spin-independent case~\cite{bg,cm}. 
Nevertheless it is useful to discuss the data that exists.

In the first resonance region, the
$\Delta(1232)$ itself gives nearly all the signal at very low $Q^2$ in the
unpolarized case and may be expected to do the same in the polarized
case.  Further, at low $Q^2$, $\Delta$ electroproduction is dominated by
the magnetic dipole amplitude~\cite{be,dm}, which leads to 
$|G_-| \approx \sqrt{3} |G_+|$  and 
\begin{equation}
A_1(\Delta,{\rm low\ } Q^2) \approx -1/2.
\end{equation}

\noindent Abe {\it et al.}\ find for $Q^2 \approx 0.5$ GeV$^2$ and in the
$\Delta$ region, that $A_1 \approx -1/3$, in qualitative agreement with
our expectation.  However, for $Q^2 \approx 1.2$ GeV$^2$, the
measured value of $A_1$ is consistent with zero (albeit also consistent
with $-1/2$ at a $2\sigma$ level).  Since there is evidence that the M1
dominance is still valid for the resonance itself, the
$A_1$ result must be due to the background and resonance giving
approximately canceling contributions.  This suggests a violation of the
strict construction of Bloom-Gilman duality, since the $Q^2$ dependence
of the resonance and background do not match.  However, $Q^2$ is still
low.  

It is reminiscent of the unpolarized case, where for the $\Delta$
Bloom-Gilman duality works (above a few GeV$^2$) in the sense of the
resonance region average matching the scaling curve, and does so without
having the resonance to continuum ratio be constant, but rather because
of an interplay between resonance and continuum~\cite{cm}.  As one falls,
the other rises,  relative to the scaling curve,  and the sum stays about
the same.  So as in the unpolarized case, the sum over channels allows
matching the scaling curve at low $Q^2$, perhaps so in the polarized case
the sum over channels will show the perturbative polarization prediction
at a lower $Q^2$ when a single channel will not.

For the second resonance region, the prediction for $A_1$ involves the
$S_{11}(1535)$ and $D_{13}(1520)$, as well as the non-resonant
background.  At very low $Q^2$, the largest resonant amplitude is the
$A_{3/2}$ for exciting the $D_{13}$~\cite{pdg}, the next largest is the
$A_{1/2}$ for the $S_{11}$~\cite{k}, and $A_{1/2}$ for the $D_{13}$ is
quite small.  However,  the resonances soon reconcile themselves to the
high $Q^2$ expectations,  as for the $D_{13}$ the $A_{1/2}$ and $A_{3/2}$
change relative size in the vicinity of 1 GeV$^2$~\cite{burkert}.   Hence,
considering the resonant contributions alone, we expect $A_1$ to be
negative at low $Q^2$ and become positive before 1 GeV$^2$.   The
available data~\cite{abe} show $A_1$ to be positive at both 0.5 and 1.2
GeV$^2$.

For polarized structure functions, in contrast to the unpolarized case,
duality must break down spectacularly at low enough $Q^2$~\cite{fc}.  The
argument goes by considering the Ellis-Jaffe integral~\cite{ej}, written
as
\begin{equation}
\int^\infty_{\nu_0} {d\nu \over \nu^2} \  g_1^p = {2 m_N \Gamma^p \over
Q^2},
\end{equation}

\noindent with $\Gamma^p$ approximately constant at high $Q^2$, and
comparing it to the Drell-Hearn-Gerasimov sum rule~\cite{dhg}, which may
be written in the form
\begin{equation}
\int^\infty_{\nu_0} {d\nu \over \nu^2} \  g_1^p(\nu,Q^2=0) = 
         - {\kappa_p^2 \over 2 m_N}.
\end{equation}
Above, $\kappa_p$ is the anomalous magnetic moment of the proton,  and
$\nu_0$ is given by the pion production threshold.  The quantity
$\Gamma_p$ is measured to be positive at $Q^2$ of several GeV$^2$ (see,
for example,~\cite{abe}).   If $g_1$ were on the average the same at very
low $Q^2$ as it is at high $Q^2$, then the right hand side of the
Drell-Hearn-Gerasimov sum rule would be positive---and it is clearly not.

We have shown that polarization structure function data in the resonance
region at higher $Q^2$ are interesting and can throw significant new
light on the issue of duality.   Facilities like Jefferson Lab, SLAC, and
HERMES can all contribute over a significant range of $Q^2$ and $W$.  
The idea that the single quark cross section sets the scale on the
average even in the resonance region gets a new field of exploration in
the polarized structure function.  Unlike the unpolarized case, one
expects its breakdown at sufficiently low $Q^2$.   Finding this breakdown
will signal the onset of a region where the final state interactions
obliterate even a remnant of perturbative physics.

We thank R. M. Davidson, Keith Griffioen, and P. Stoler for useful
comments, and members of the Jefferson Lab Hall A collaboration for
stimulating discussions.  CEC thanks the NSF for support under grant
PHY-9600415; and NCM is grateful to the U. S. Department of Energy for
its support through grant DE-FG02-88ER40448.

\end{document}